\documentclass[preprint,prc,showpacs,preprintnumbers,
               superscriptaddress,floatfix]{revtex4}

\usepackage{graphicx,latexsym,amssymb}%, epsfig}
\usepackage{multirow,amsmath,array,booktabs}
\usepackage{subfigure}
\usepackage[figuresright]{rotating}

\newcommand{\ben}{\begin{enumerate}}
\newcommand{\een}{\end{enumerate}}
\newcommand{\beq}{\begin{equation}}
\newcommand{\eeq}{\end{equation}}
\newcommand{\beqn}{\begin{eqnarray}}
\newcommand{\eeqn}{\end{eqnarray}}

\newcommand{\beqd}{\begin{eqnarray*}}
\newcommand{\eeqd}{\end{eqnarray*}}
\newcommand{\bea}{\begin{array}}
\newcommand{\eea}{\end{array}}
\newcommand{\bcen}{\begin{center}}
\newcommand{\ecen}{\end{center}}
\newcommand{\btab}{\begin{tabular}}
\newcommand{\etab}{\end{tabular}}
\newcommand{\bsub}{\begin{subequations}}
\newcommand{\esub}{\end{subequations}}
\newcommand{\bit}{\begin{itemize}}
\newcommand{\eit}{\end{itemize}}
\newcommand{\brule}{\begin{ruledtabular}}
\newcommand{\erule}{\end{ruledtabular}}
\newcommand{\bpm}{\begin{pmatrix}}
\newcommand{\epm}{\end{pmatrix}}
\newcommand{\lc}{\left<}
\newcommand{\rc}{\right>}
\newcommand{\lr}{\left|}
\newcommand{\rl}{\right|}

\newcommand{\Lb}{\left\{}
\newcommand{\Rb}{\right\}}

\newcommand{\bay}{\begin{array}}
\newcommand{\eay}{\end{array}}

\newcommand{\balp}{\mbox{\boldmath$\alpha$}}

\newcommand{\br}{{\mathbf{r}}}

\begin{document}

\title{\vspace{1cm} Multi-chiral doublets in one single nucleus}

\author{J. Meng}\thanks{e-mail: mengj@pku.edu.cn}
\affiliation{School of Physics, Peking University, Beijing 100871}
\affiliation{Institute of Theoretical Physics, Chinese Academy of
Science, Beijing 100080}
\affiliation{Center of Theoretical Nuclear Physics, National Laboratory of \\
       Heavy Ion Accelerator, Lanzhou 730000}
\author{J. Peng}
\affiliation{School of Physics, Peking University, Beijing 100871}
\author{S. Q. Zhang}
\affiliation{School of Physics, Peking University, Beijing 100871}
\author{S.-G. Zhou}
\affiliation{Institute of Theoretical Physics, Chinese Academy of
Science, Beijing 100080}
\affiliation{Center of Theoretical Nuclear Physics, National Laboratory of \\
       Heavy Ion Accelerator, Lanzhou 730000}

\begin{abstract}

Adiabatic and configuration-fixed constraint triaxial relativistic
mean field (RMF) approaches are developed for the first time and a
new phenomenon, the existence of multi-chiral doublets (M$\chi$D),
i.e., more than one pairs of chiral doublets bands in one single
nucleus, is suggested for $^{106}$Rh based on the triaxial
deformations together with their corresponding proton and neutron
configurations.

\end{abstract}

\pacs{21.10.Re, 21.60.Jz, 21.10.Pc, 21.10.Gv, 27.60.+j}

\maketitle

%\section{Introduction}

The handedness or chirality is a subject of general interests in
molecular physics, elementary particles, and optical physics. The
occurrence of chirality in nuclear physics was suggested in
1997~\cite{Frauendorf97} and the predicted patterns of spectra
exhibiting chirality, i.e., the chiral doublets bands, were
experimentally observed in 2001~\cite{Starosta01a}.

Since the pioneer work on chirality in nuclear physics, lots of
efforts have been made to understand the new phenomena and explore
their possible existence in the nuclear chart, e.g.,
~\cite{Dimitrov00, Peng03a, Balabanski, Olbratowski04,
Koike04,Vaman04,Joshi04,Timar04}. Experimentally, the chiral
doublets bands have been identified in many nuclei in $A \sim 130$
mass region with the earlier suggested configuration $\pi h_{11/2}
\otimes \nu h^{-1}_{11/2}$, $A \sim 100$ with $\pi
g^{-1}_{9/2}\otimes\nu h_{11/2}$, and $A \sim 190$ with $\pi
h_{9/2}\otimes\nu i^{-1}_{13/2}$. On theoretical aspect, the
chiral symmetry breaking was firstly predicted in the
particle-rotor model (PRM) and tilted axis cranking (TAC) approach
for triaxially deformed nuclei~\cite{Frauendorf97}. It has been
investigated later in hybrid Woods-Saxon and Nilsson model
combined with shell correction method~\cite{Dimitrov00} as well as
the Skyrme-Hartree-Fock cranking approach~\cite{Olbratowski04},
and its selection rules for electromagnetic transitions have been
discussed in a simple PRM~\cite{Koike04}.

For triaxially deformed rotational nucleus, the collective angular
momentum favors alignment with the intermediate axis, which in
this case has the largest moment of inertia. Meanwhile, the
valence particle and hole angular momentum vectors align along the
nuclear short and long axis, respectively. These orientations
maximize the overlap of the particle densities with the triaxial
core and minimize the interaction energy. The three mutually
perpendicular angular momenta can be arranged to form two systems
with opposite chirality, a left- and a right-handedness. They are
transformed into each other by the chiral operator which combines
time reversal and spacial rotation of 180$^\circ$,
$\chi={\cal{TR}}(\pi)$. The breaking of the chiral symmetry in
atomic nucleus is observed due to the quantum tunnelling between
the systems with opposite chirality.

The description for the quantum tunnelling of the chiral partners
is beyond the mean field, as the usual cranking approach is a
semiclassical model and the total angular momentum is not a good
quantum number~\cite{Frauendorf01}. In contrast, the PRM is better
suited for this purpose, which however is so far confined for one
particle and one hole configurations only~\cite{Frauendorf97,
Peng03a, Koike04}. Its generalization for more particles and/or
holes are still under development. Furthermore the deformation
$\gamma$ and configurations in PRM are not self-consistent rather
as inputs of the model.

The relativistic mean field (RMF) theory has received wide
attention due to its success in describing the properties of
nuclei and many nuclear phenomena for the past
years~\cite{Ring96,Meng98npa}. It is interesting to search for
nuclei with triaxial deformation and configurations with not only
one particle and one hole but also multi-particle and multi-hole
suitable for chirality in RMF theory. A multi-dimensional
microscopic cranking RMF model is very time consuming and has only
been applied in magnetic rotation so far~\cite{Madokoro00}.

In this Letter, we will develop the adiabatic and
configuration-fixed constraint triaxial RMF approaches to
investigate the triaxial shape coexistence and the possible chiral
doublets bands in $A\sim100$ mass region. The existence of
multi-chiral doublets (M$\chi$D), i.e., more than one pairs of
chiral doublets bands in one single nucleus, will be suggested via
the examining the deformation and the corresponding
configurations.

%\section{Formulation}

In RMF theory, the nuclei are characterized by an attactive scalar
field $S(\br)$ and a repulsive vector field $V(\br)$ in the Dirac
equation,
   \beq
     \{ -i\balp\cdot {\mbox \boldmath \nabla}
         + V(\br) + \beta [ M +S(\br) ] \} \psi_i
      =\ \varepsilon_i\psi_i,
   \eeq
which can be solved by expanding separately the upper and lower
components of the spinor $\psi_i$ in terms of eigenfunctions of
the three-dimensional deformed oscillator in Cartesian coordinates
$\phi_{\alpha}(\br)$ and its time reversal state
$\phi_{\bar{\tilde{\alpha}}}(\br)=\hat{T} \phi_{\alpha}(\br)$ with
$\hat{T}= i \hat{\sigma}_y K$ ,
   \beq
     \psi_i(\br)
      =  \left(\begin{array}{c}
           \sum\limits_{\alpha}f^{(i)}_{\alpha}\phi_{\alpha}(\br) \\
           \sum\limits_{\tilde{\alpha}}ig^{(i)}_{\tilde{\alpha}}\phi_{\bar{\tilde{\alpha}}}(\br)\\
         \end{array} \right).
   \eeq
In order to avoid the complex matrix diagonaliztion problem in
Ref.~\cite{Hirata96}, $\phi_{\alpha}(\br)$ is written as a product
of three Hermite polynomials~\cite{Koepf89},
   \beq
      \phi_\alpha(\br)=\frac{i^{n_y}}{\sqrt{2}}\phi_{n_x}(x)\phi_{n_y}(y)\phi_{n_z}(z)
               \left(\begin{array}{c}1\\
               (-1)^{n_x+1}\end{array}\right).
   \eeq
The meson fields are expanded similarly with the same deformation
$\beta_0$ and $\gamma_0$ for the basis, but an oscillator length
smaller by a factor of $\sqrt{2}$ than that for the nucleons
($b_B=b_0/\sqrt{2}$) has been used in order to simplify the
calculations and to avoid the necessity of additional parameters.
The oscillator frequency is $\hbar\omega_0=41 A^{-1/3}$ MeV.

\begin{figure}[htbp]
 \centering
 \includegraphics[width=8.0cm]{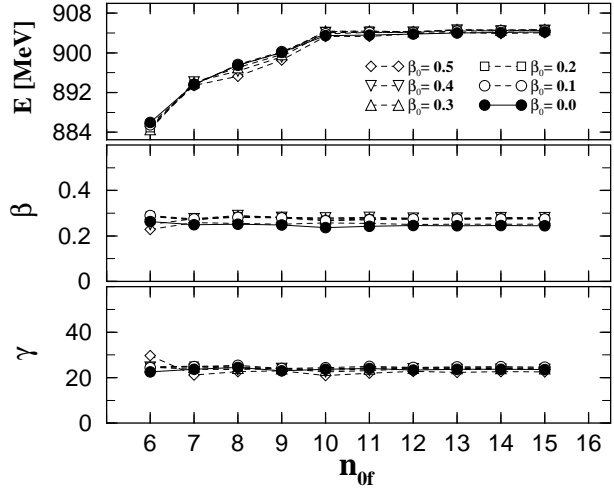}
 \caption{Total binding energy, deformation $\beta$ and $\gamma$
  calculated with PK1 for $^{106}$Rh
  as functions of the number of expanded oscillator shells for
  fermions n$_{\rm{0f}}$.
  The filled circles, open circles, squares, triangle ups,
  triangle downs, and diamonds represent the results
  as the deformation of basis $\beta_0$ equals
  to 0.0, 0.1, 0.2, 0.3, 0.4 and 0.5, respectively. }
 \label{fig:base}
\end{figure}

In order to check the convergence of the results with the number
of expanded oscillator shells for fermions n$_{\rm{0f}}$, the
binding energies (upper) and the deformations $\beta$ (middle) and
$\gamma$ (lower) calculated with effective interaction
PK1~\cite{Long04} for $^{106}$Rh as functions of n$_{\rm{0f}}$ are
presented in Fig.~\ref{fig:base}. The legends (filled circles,
open circles, squares, triangle ups, triangle downs, and diamonds)
represent the results obtained in basis with different deformation
($\beta_0$= 0.0, 0.1, 0.2, 0.3, 0.4 and 0.5, respectively). It
shows that as long as n$_{\rm{0f}} \geq 10$, the binding energies
as well as the deformations $\beta$ and $\gamma$ obtained are
almost the same. Furthermore they are independent on the
deformation $\beta_0$ of the basis. Similar convergence check for
the bosons has also been done. In the following, a spherical basis
with 12 major oscillator shells for fermions and 10 shells for
bosons will be used, which gives an error less than 0.1$\%$ for
the binding energy.

%\section{Results and Discussion}

In general the triaxial RMF calculation leads to only some local
minima. In order to get the ground state for triaxial deformed
nucleus, constraint calculations are necessary and in principal
such calculations should be carried out in two-dimensional $\beta$
and $\gamma$ plane. As such two-dimensional constraint
calculations turn out to be expensive even for modern computer
facilities, therefore the constraint calculations with $\lc
\hat{Q}_{20}^2+2\hat{Q}_{22}^2 \rc$, i.e., $\beta^2$, are carried
out to search for the ground state for triaxially deformed
nucleus.

\begin{figure}[tbp]
 \centering
 \includegraphics[width=6.0cm]{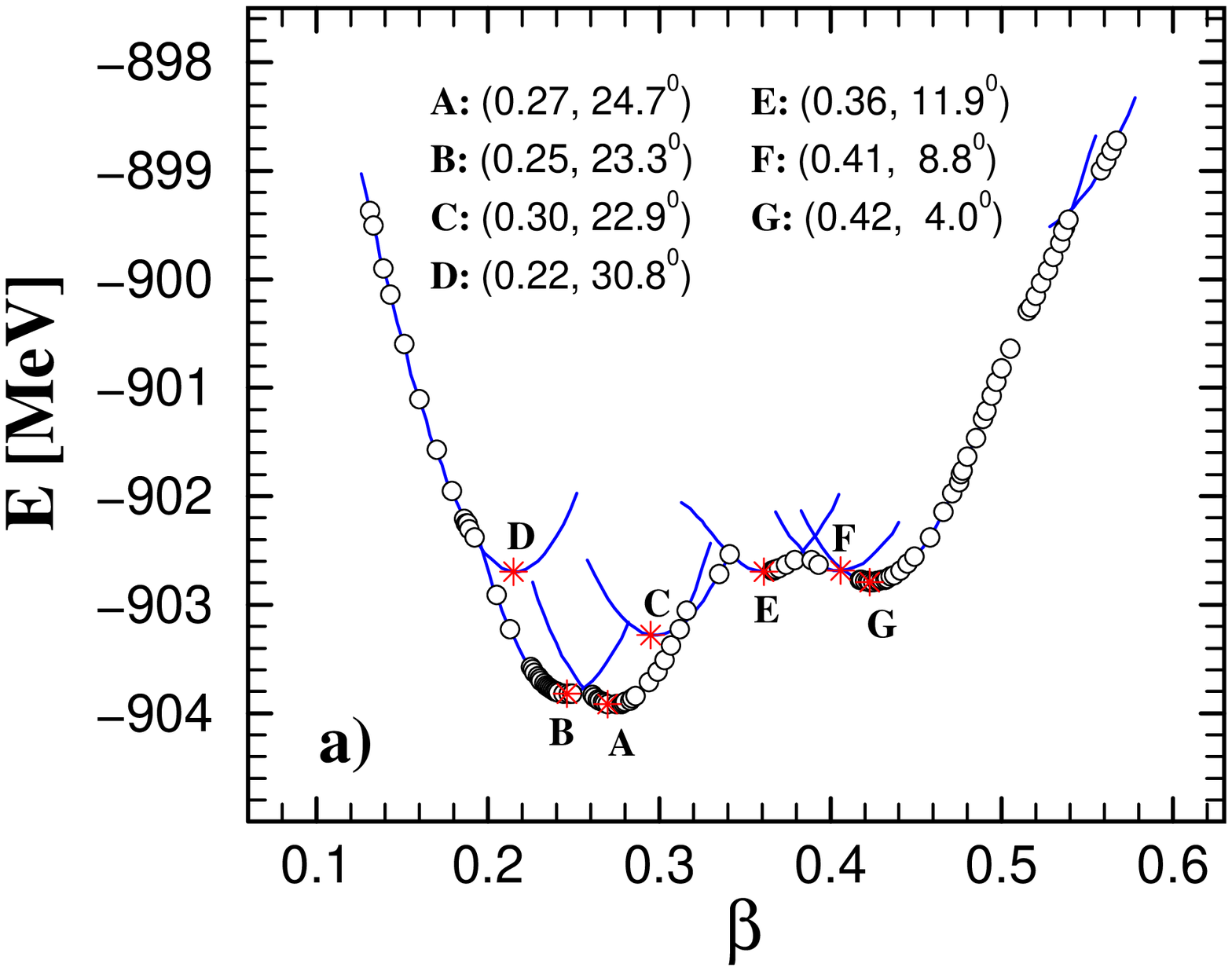}
 \includegraphics[width=6.0cm]{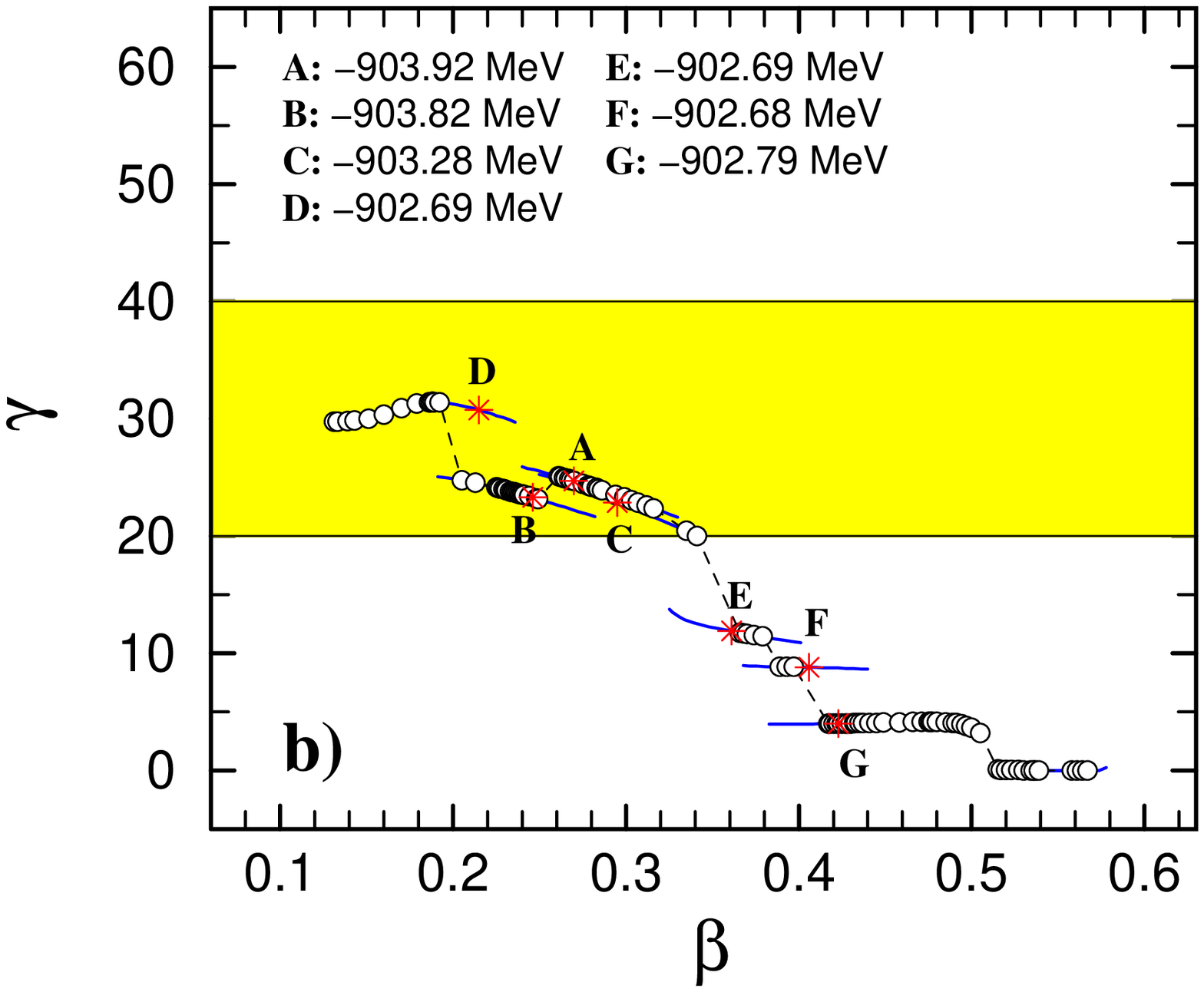}
  \caption{(color online)
    The energy surfaces (a) and the deformations $\gamma$ (b)
    as functions of deformation $\beta$ in adiabatic
    (open circles) and configuration-fixed (solid lines)
    constraint triaxial RMF calculation with PK1 for $^{106}$Rh.
    The minima in the energy
    surfaces are represented as stars and labelled respectively
    as A, B, C, D, E, F and G. Their corresponding deformations
    $\beta$ and $\gamma$ together with their energies are
    respectively given in (a) and (b).  }
 \label{fig:Rh106}
\end{figure}

The energy surface and the deformation $\gamma$ as functions of
deformation $\beta$ in adiabatic constraint triaxial RMF
calculation with PK1 for $^{106}$Rh are presented as open circles
in Fig.~\ref{fig:Rh106}(a) and (b) respectively. There are some
irregularities in the energy surface. Furthermore some local
minima are too obscure to be recognized and it is technically
difficult to understand their corresponding single particle
configurations. We therefore performed the configuration-fixed
constraint calculation similar to what have been done in the
non-relativistic case~\cite{GuoL04}. Starting from any point in
the energy surface in the adiabatic constraint calculations, the
configuration-fixed constraint calculation requires that the
single particle orbits occupied are fixed during the constraint
calculation, i.e.,
   \beq
      \lr\lc \psi_j(\beta+\delta\beta)
           | \psi_i(\beta)\rc\rl
      \approx 1.
   \eeq
The energy surfaces and the deformations $\gamma$ in
configuration-fixed calculations with PK1 for $^{106}$Rh are given
as solid lines in Fig.~\ref{fig:Rh106}(a) and (b) respectively.
For each fixed configuration, the constraint calculation gives a
continuous and smooth curve for the energy surface and the
deformation $\gamma$ as a function of deformation $\beta$. The
irregularities in the adiabatic energy surface disappear. In
comparison, the minima in the energy surfaces of the
configuration-fixed constraint calculations become obvious, which
are respectively represented by stars and labelled as A, B, C, D,
E, F and G. Their corresponding deformations $\beta$ and $\gamma$
together with their binding energies are respectively given in
Fig.~\ref{fig:Rh106}(a) and (b). It is interesting to note that,
for each fixed configuration, the deformation $\gamma$ is
approximately a constant (as in Fig~\ref{fig:Rh106}(b)), which
means that the deformation $\gamma$ is mainly determined by its
corresponding configuration.

The energies for these minima including the ground state are
within 1.3 MeV to each other but correspond to different
deformations $\beta$ and $\gamma$, which is a good example of the
shape coexistence. The shape coexistence here is different from
the spherical, oblate and prolate shape coexistence, e.g.,  in
neutron-deficient Pt, Hg and Pb isotopes. It is the triaxial shape
coexistence, i.e., for the ground state A: the binding energy $E$=
903.92 MeV (the data 906.72 MeV), deformation $\beta= 0.27$ and
$\gamma= 24.7^{~\circ}$, the excited minima B: $E$= 903.82 MeV,
$\beta=0.25$ and $\gamma=23.3^{~\circ}$, C: $E$= 903.28 MeV,
$\beta=0.30$ and $\gamma=22.9^{~\circ}$, and D: $E$= 902.69 MeV,
$\beta=0.22$, $\gamma=30.8^{~\circ}$. All the states A, B, C, and
D have deformation $\beta$ and $\gamma$ suitable for
chirality~\cite{Frauendorf97,Peng03a}. As these states are all in
one single nucleus, if chiral doublets bands can be built on these
states, it may lead to a new phenomenon, the existence of M$\chi$D
in one single nucleus. Therefore, in the following, we will
investigate their proton and neutron configurations in detail to
see whether the particle and hole configurations required by
chirality are available.

\begin{figure}[tbp]
 \centering
 \includegraphics[width=6.0cm]{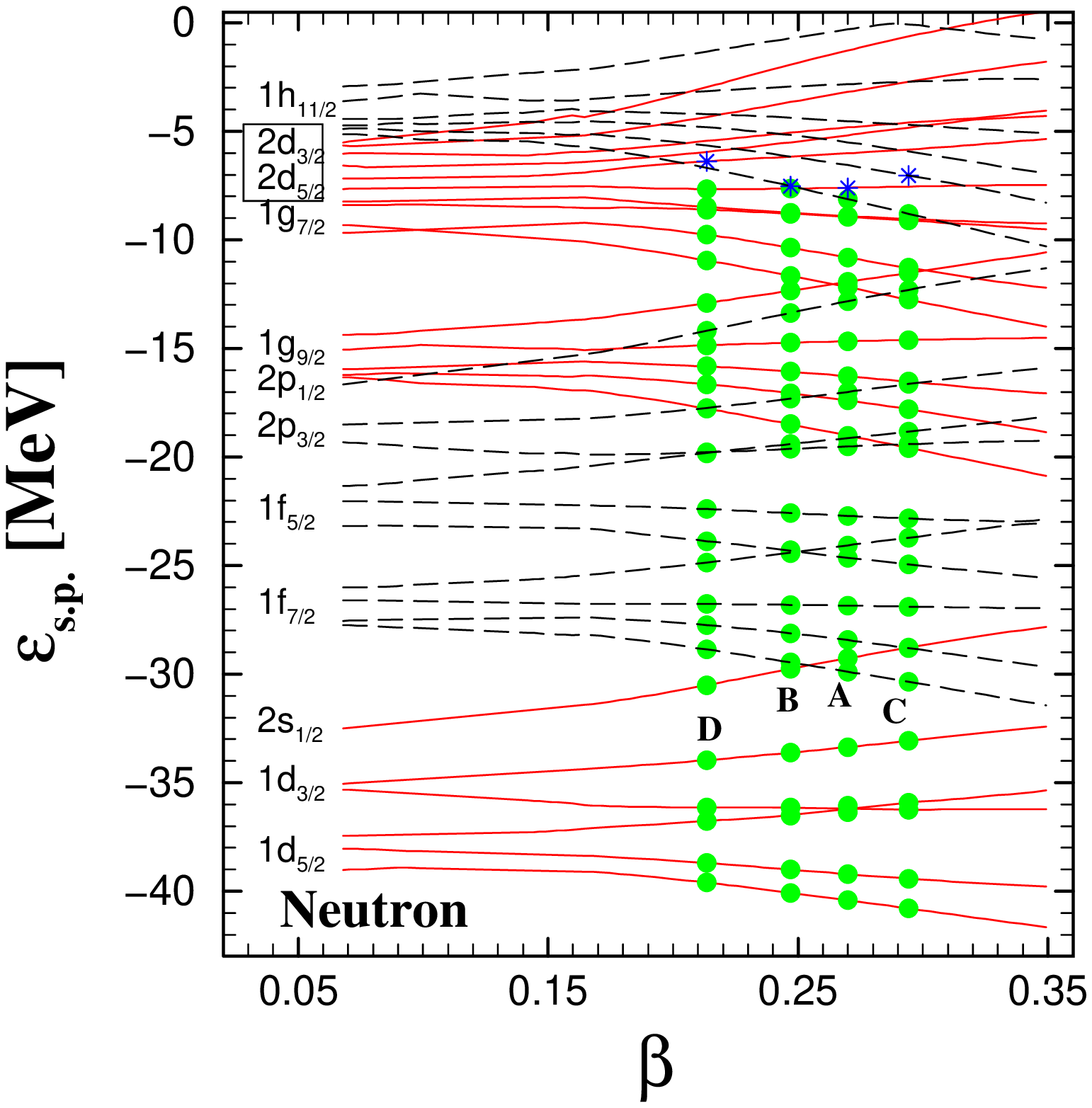}
 \includegraphics[width=6.0cm]{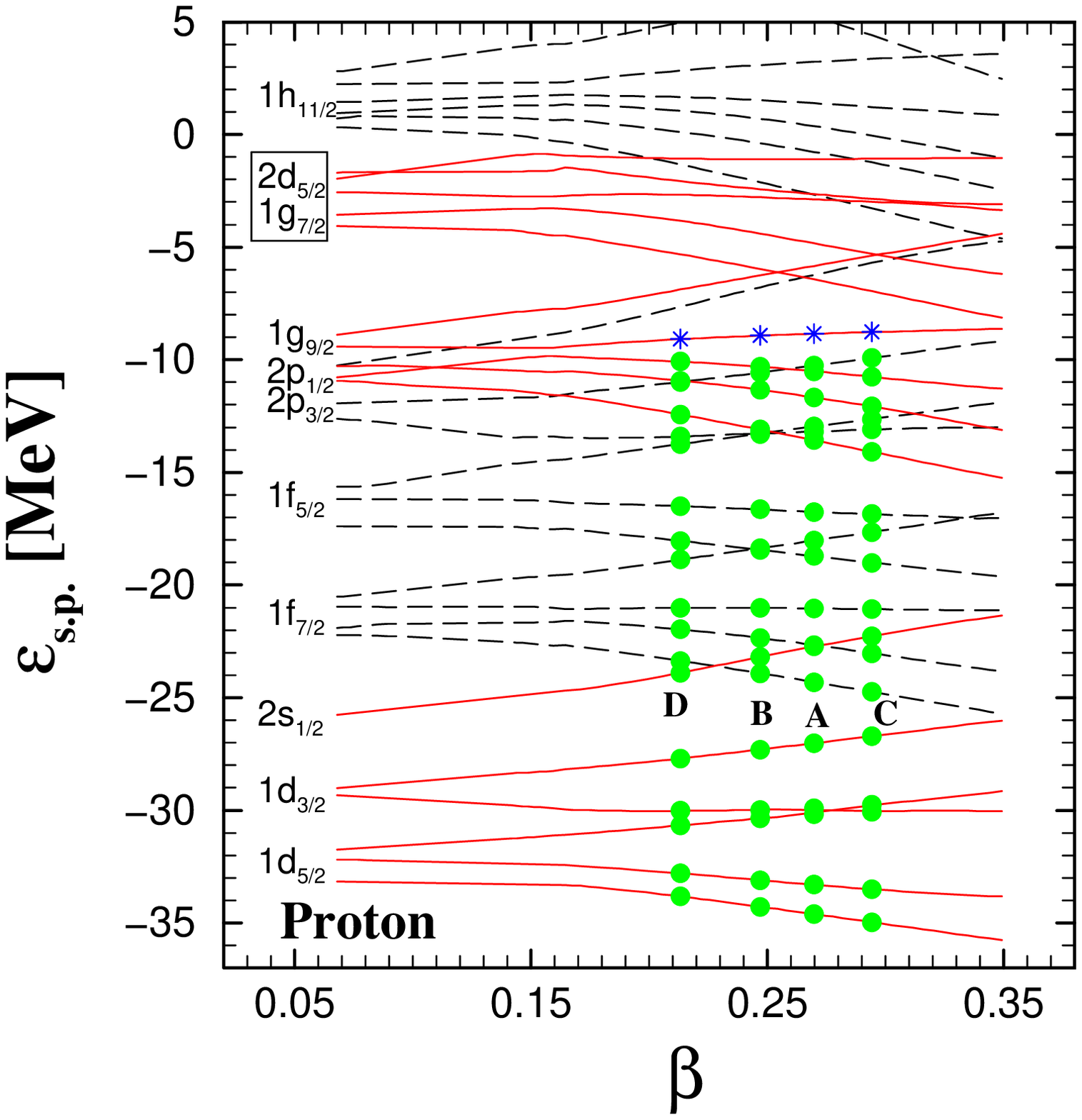}
 \caption{(color online)
 The neutron and proton single particle levels
 obtained in configuration-fixed
 constraint triaxial RMF calculations with PK1 for $^{106}$Rh
 as functions of the deformation $\beta$. Positive (negative)
 parity states are marked by solid (dashed) lines.
 The occupations corresponding to the
 minima in Fig.~\ref{fig:Rh106} are represented by filled
 circles (two particles) and stars (one particle).
 }
 \label{fig:Rh106orbit}
\end{figure}

Performing the configuration-fixed constraint calculations for the
ground state, the single particle levels can be obtained as
functions of deformation $\beta$. The difference in single
particle levels obtained by choosing other minima is negligible.
The neutron and proton single particle levels obtained in such a
way are presented in Fig.~\ref{fig:Rh106orbit}. The positive
(negative) parity states are marked by solid (dashed) lines and
the occupations corresponding to the minima in
Fig.~\ref{fig:Rh106} are represented by filled circles (two
particles) and stars (one particle). The corresponding quantum
numbers for spherical case are labelled at the left side of the
levels. As the energy curves become very stiff for small
deformation in Fig.~\ref{fig:Rh106}, the present
configuration-fixed constraint calculations cannot be performed
for deformation $\beta < 0.06$. Therefore we cannot distinguish
the occupation of some low-$j$ orbits, e.g., the last occupied
neutron orbit for state A may come from $2d_{5/2}$ or $2d_{3/2}$,
as marked in the figures. However, as the nuclear chirality is
essentially determined by the high-$j$ orbits, such illegibility
does not influence the conclusions here.

The binding energies, deformations $\beta$ and $\gamma$ as well as
the corresponding configurations extracted from
Fig.~\ref{fig:Rh106orbit} for the minima A, B, C, and D in
$^{106}$Rh obtained in the configuration-fixed constraint triaxial
RMF calculations with PK1 are listed in Table.~\ref{tab:occupy}.
Except the state D, in which there is no high-$j$ neutron valence
particle, we found the high-$j$ proton and neutron configuration
for the ground state (state A) as
 $\pi(1g_{9/2})^{-3} \otimes \nu (1h_{11/2})^2 $,
 state B as
 $\pi(1g_{9/2})^{-3} \otimes \nu (1h_{11/2})^1 $,
 and state C as
 $\pi(1g_{9/2})^{-3} \otimes \nu(1h_{11/2})^3$,
respectively. All of them have high-$j$ proton holes and high-$j$
neutron particles configurations, which together with their
triaxial deformations favor the construction of the chiral
doublets bands. It is interesting to note that the states A and B
compete strongly with each other in energy. However, due to
different parities, the states A and B do not mix up and are
possible to produce the new phenomenon M$\chi$D. So far, a pair of
chiral doublets bands with negative parity have been observed in
$^{106}$Rh~\cite{Joshi04}. It will be interesting to search for
other chiral doublets bands in this nucleus.

\begin{figure}[tbp]
 \centering
 \includegraphics[width=8.0cm]{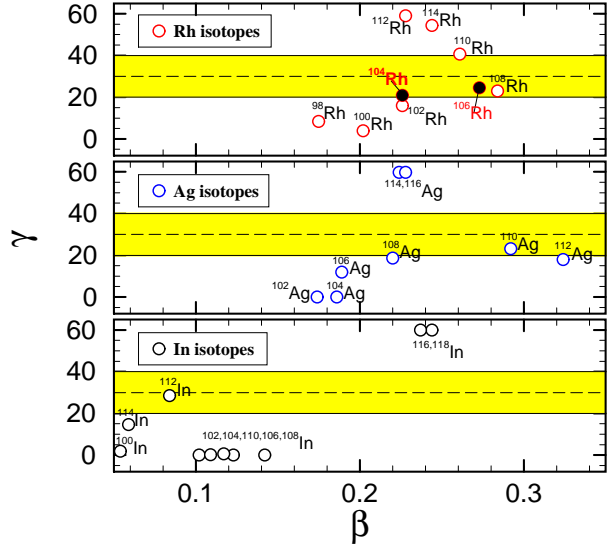}
 \caption{(color online)
 The deformations $\beta$ and $\gamma$ for the ground states
 in Rh (upper), Ag (middle) and In (lower) isotopes in
 constraint triaxial RMF calculations with PK1.
 The shaded area represents the favorable deformation $\gamma$
 for nuclear chirality.
 The nuclei $^{104}$Rh and $^{106}$Rh, in which the chiral
 doublets bands have been observed, are marked as
 filled circles.}
 \label{fig:betagam}
\end{figure}

Similar to $^{106}$Rh, detailed constraint calculations are also
performed for other isotopes in $A\sim100$ mass region and the
possibilities of M$\chi$D exist in other nuclei as well. The
results will be published elsewhere in details. Here in
Fig.~\ref{fig:betagam}, the deformations $\beta$ and $\gamma$ for
ground states in $^{98-114}$Rh, $^{102-116}$Ag and $^{100-118}$In
isotopes are presented. The shaded area represents the favorable
deformation $\gamma$ for nuclear chirality. The nuclei, $^{104}$Rh
and $^{106}$Rh, in which the chiral doublets bands have been
observed~\cite{Vaman04,Joshi04}, are marked as filled circles.
Apart from $^{104}$Rh and $^{106}$Rh, the favorable deformation
$\gamma$ for chirality has been found in $^{102,108,110}$Rh,
$^{108-112}$Ag and $^{112}$In, which implicates that more chiral
doublets bands can be expected in $A\sim100$ mass region. For each
isotope chain, the deformation $\gamma$ varies with the neutron
number due to different occupation in $\nu h_{11/2}$.

%\section{Summary}

In summary, adiabatic and configuration-fixed constraint triaxial
RMF approaches are developed for the first time to investigate the
triaxial shape coexistence and possible chiral doublets bands.  A
new phenomenon, the existence of multi-chiral doublets (M$\chi$D),
i.e., more than one pairs of chiral doublets bands in one single
nucleus, is suggested in $^{106}$Rh by examining the deformation
and the corresponding configurations. Similar investigations have
also been done for Rh, Ag and In isotopes and also with other
effective interactions. The observation of M$\chi$D is very
promising in this mass region.

\vspace{2em}This work is partly supported by the Major State Basic
Research Development Program Under Contract Number G2000077407,
the National Natural Science Foundation of China under Grant No.
10435010, 10221003, and 10475003,  the Doctoral Program Foundation
from the Ministry of Education in China, and the Knowledge
Innovation Project of Chinese Academy of Sciences under contract
Number KJCX2-SW-N02.

%####################################################################

%###########Figs and tables###############
 \begin{table}[!h]
 \centering \caption{The binding energies,
   deformations $\beta$ and $\gamma$ and
   the corresponding configurations for the minima
   A, B, C, and D in $^{106}$Rh obtained in the
   configuration-fixed constraint triaxial RMF
   calculations with PK1.}
 \label{tab:occupy}
 \btab{ |c| c| c| c|c|  }\hline\hline
       State &~E~(MeV)~~&~~~$\beta$~~~&~~~$\gamma$~~~&~~~~~Configurations \\ \hline
      A      & 903.9150 & 0.270 & 24.7$^\circ$ &  $\pi(1g_{9/2})^{-3} \otimes \nu \Lb (1h_{11/2})^2 [(2d_{5/2})^1\right. ~{\rm or} ~  \left.(2d_{3/2})^1]\Rb$\\ \hline
      B~~&~~~903.8196~~~&~~0.246~~&~~23.3$^\circ$~~&~$\pi(1g_{9/2})^{-3}\otimes \nu \Lb (1h_{11/2})^1 [(2d_{5/2})^2\right.$~or~$\left.(2d_{3/2})^2]~~\Rb$\\ \hline
      C~~&~~~903.2790~~~&~~0.295~~&~~22.9$^\circ$~~& ~~~~$\pi(1g_{9/2})^{-3}\otimes \nu(1h_{11/2})^3$ \\\hline
      D~~&~~~902.6960~~~&~~0.215~~&~~30.8$^\circ$~~& ~~~  $\pi(1g_{9/2})^{-3}\otimes \nu[(2d_{5/2})^3$~or~$(2d_{3/2})^3]$\\ \hline
             \hline\etab
 \end{table}

\end{document}